\documentstyle[12pt]{article}
\newcommand {\m}{\mu}
\newcommand {\n}{\nu}
\newcommand {\pl}{\partial}
\newcommand {\p} {\phi}

\newcommand {\al}{\alpha}

\newcommand {\ga}{\gamma}

\newcommand {\la}{\lambda}
\newcommand {\La}{\Lambda}
\newcommand {\si}{\sigma}
\newcommand {\th}{\theta}

\newcommand {\ep}{\epsilon}

\newcommand {\na}{\nabla}
\newcommand {\del}  {\delta}

\newcommand {\mn}{{\mu\nu}}
\newcommand {\ls}   {{\lambda\sigma}}

\newcommand {\half}{ {\frac{1}{2}} }

\newcommand {\sqg} {\sqrt{g}}

\newcommand {\Lcal}{{\cal L}}

\newcommand {\Dcal}{{\cal D}}
\newcommand {\Dvec}{{\vec D}}

\newcommand {\Vvec}{{\vec V}}

\newcommand {\ra} {\rightarrow}
\newcommand {\pr}   {{\quad .}}
\newcommand {\com}  {{\quad ,}}
\newcommand {\q}    {\quad}

\newcommand {\nn}    {\nonumber}
\newcommand {\vs}[1]  { \vspace*{#1 cm} }
\newcounter{eq}
\newcounter{sc}

\newcommand {\MPL}  {Mod.Phys.Lett.}
\newcommand {\NP}   {Nucl.Phys.}
\newcommand {\PL}   {Phys.Lett.}
\newcommand {\PR}   {Phys.Rev.}
\newcommand {\PRL}   {Phys.Rev.Lett.}
\newcommand {\CMP}  {Commun.Math.Phys.}

\newcommand {\CQG}  {Class.Quantum.Grav.}
\def\overleftrightarrow#1{\vbox{\ialign{##\crcr
 $\leftrightarrow$\crcr\noalign{\kern-1pt\nointerlineskip}
 $\hfil\displaystyle{#1}\hfil$\crcr}}}

\def\barpsi{{\bar\psi}}

\def\dslash{{}\hbox{\hskip2pt\vtop
 {\baselineskip23pt\hbox{}\vskip-24pt\hbox{/}}
 \hskip-11.5pt $\partial$}}
\def\Dslash{{}\hbox{\hskip2pt\vtop
 {\baselineskip23pt\hbox{}\vskip-24pt\hbox{/}}
 \hskip-11.5pt $D$}}

\def\gsym{\delta}

\newlength{\minitwocolumn}
\begin{document}

%%%%%%%%%%%%%%%%%%%%%%%%%%%%%%%%%%%%%%%%%%%%%%%%%%%%%%%%%%%%%%%%%%
%%%%%%%%%%%%%%%%%%%%%%%% Title %%%%%%%%%%%%%%%%%%%%%%%%%%%%%%%%%%%
%%%%%%%%%%%%%%%%%%%%%%%%%%%%%%%%%%%%%%%%%%%%%%%%%%%%%%%%%%%%%%%%%%

\begin{flushright}
DAMTP/96-87\\
Oct.,1996\\
hep-th/9610136
\end{flushright}
\vspace{24pt}

%\magnification=\magstep1
\pagestyle{empty}
\baselineskip15pt
%\font\cmssB=cmss17
%\font\cmssS=cmss10

\begin{center}
{\large\bf Gauge Symmetry of the Heat-Kernel and \\
Anomaly Formulas \vskip 1mm
}

\vspace{10mm}

Shoichi ICHINOSE
          \footnote{
          On leave of absence, until Jan. 31,1997,  from
          Department of Physics, University of Shizuoka,
          Yada 52-1, Shizuoka 422, Japan.
          E-mail address:\ s.ichinose@damtp.cam.ac.uk\ ;\
          ichinose@momo1.u-shizuoka-ken.ac.jp
                  }
and Noriaki IKEDA${}^{\dag}$ 
\footnote
{ E-mail address:\ nori@kurims.kyoto-u.ac.jp }
\\
\vspace{5mm}
DAMTP, University of Cambridge,       \\
Silver Street, Cambridge CB3 9EW, UK                  \\
${}^{\dag}$Research Institute for Mathematical Sciences \\
Kyoto University, Kyoto 606-01, Japan
\end{center}
%\date{July, 1996}

%\maketitle

\vspace{15mm}
\begin{abstract}
We consider a gauge symmetry in a quantum Hilbert space. The
symmetry leads to that of the heat-kernel and of
the anomaly formulae which were previously obtained by the authors. 
This greatly simplifies and clarifies the structure of the
formulae.
We explicitly obtain the anomaly formulae 
in two and four dimensions, which ``unify''
all kinds of anomaly.
The symmetry corresponds to that of the counterterm formulae in the
background field method. As an example, 
the non-abelian anomaly is considered.
\end{abstract}

\newpage
\pagestyle{plain}
\pagenumbering{arabic}
%\setcounter{page}{1}

%%%%%%%%%%%%%%%%%%%%%%%%%%%%%%%%%%%%%%%%%%%%%%%%%%%%%%%%%%%%%%%%%%
%%%%%%%%%%%%%%%%%%%%%%%% Article %%%%%%%%%%%%%%%%%%%%%%%%%%%%%%%%%
%%%%%%%%%%%%%%%%%%%%%%%%%%%%%%%%%%%%%%%%%%%%%%%%%%%%%%%%%%%%%%%%%%

\rm
%%%%%%%%%%%%%%%%%%%%%%%%%%%%%%%%%%%%%%%%%%%%%%%%%%%%%%%%%%%%%%%%%%%%%
%%%%%%%%%%%%%%%%%%%%%%%%%%%%%%   SEC  1    %%%%%%%%%%%%%%%%%%%%%%%%%%
%%%%%%%%%%%%%%%%%%%%%%%%%%%%%%%%%%%%%%%%%%%%%%%%%%%%%%%%%%%%%%%%%%%%%
\section{Introduction}

Like renormalization, anomalies are an important
aspect of the quantum field theory(QFT) and have been provided us
with important insights. There are many kinds of anomalies (Weyl,
U(1) chiral, Non-abelian, gravitational U(1) chiral, pure
gravitational (Einstein), etc.) in different theories and in
different dimensions. The applications are diverse and
range  from a
phenomenological one to a purely theoretical or mathematical one.
In this circumstance, it is useful,
from both formal and practical viewpoints,  to look at all anomalies 
from one framework. With this in mind, we present anomaly
formulas, which are valid for all kinds of anomalies of all theories
in 2 and 4 spacetime dimensions.

In the case of renormalization, 't Hooft\cite{tH} succeeded in
determining the 1-loop counter-term formula, which is valid for
all 4-dim theories. 
\footnote{
It was generalized to higher-loops
\cite{Abbott81,IO82}
and to higher-dimensions\cite{deven85}.
          }
It was exploited, from mid to late 70's, for the (un)renormalizability
check of various gravity and supergravity theories. On the other
hand, the anomaly, which has the same origin (ultraviolet 
divergences of local interaction),
has not been explored as much. Important progress in a
``unified treatment'' of anomalies was made by
Fujikawa\cite{F79} who noted that all anomalies
can be interpreted in the path-integral formalism as a change
in the measure due to a symmetry transformation. We take this standpoint
and realize the ``unified'' approach in a more manifest way as
't Hooft did with renormalization\cite{II}.

Anomalies are grouped into two types:\ those which are
related to the scale transformation (i.e. the Weyl anomaly),
and those related to
the chiral transformation (that is the chiral anomaly). 
The former is caused by
the regularization of the unit operator $1\cdot\del^n(x-y)$, 
while the latter is
by the regularization of the chiral operator $\ga_5\cdot\del^n(x-y)$.
%\cite{Ftext93}.
We know of many regularization methods in QFT
and choose an appropriate one depending on the specific
problem. 
In the case of anomalies, popular choices are regularizations of 
Pauli-Villars, heat-kernel
and higher-derivative method.
(Dimensional regularization is not appropriate for chiral anomalies due
to the difficulty in treating $\ga_5$).
In this paper we use heat-kernel regularization. Its characteristics
are as follows: 1) we can formulate the theory explicitly in a
covariant way even though a dimensional parameter, $t$, called
{\it proper-time} is introduced\cite{S51}. This is due to
the fact that the parameter is introduced not as a mass
but as an additional component of the spacetime
(time or temperature). This must be
compared with the case of Pauli-Villars;\ 
2) A key quantity $G(x,y;t)$\ , called the
{\it heat-kernel}, is defined by a heat equation with an
elliptic differential operator. It has been closely examined 
and is well-defined mathematically\cite{Sak,Gil75,Gil2}. 
This point can be compared with the higher-derivative
regularization. 
3)This regularization gives a similar result to the dimensional
regularization. Dimensional regularization essentially picks up only
logarithmic divergences, which encapsulate the most important information
of the QFT. The same information appears, in the heat-kernel,
as  a $t^0$-order term which is the remaining finite term in the
limit $t\ra +0$. 

The symmetry of the heat kernel is derived
from a gauge symmetry of the quantum Hilbert space on which
the elliptic differential operator acts.
The formulae of anomalies are shown to satisfy the gauge
symmetry. The symmetry clarifies the anomaly formulas and simplifies the
calculation.
The formula is very powerful as in the
case of the counterterm formula of 'tHooft\cite{tH}. 
The 4-dim anomaly formula coincides with the 't Hooft's
1-loop counterterm formula up to total derivative terms.
However, contrary to the case of counterterms,
the total derivative terms have significance in the anomaly calculation.

In this paper, we take the non-abelian anomaly in a 4-dim chiral
gauge model as a concrete application of the anomaly formulas.
We calculate both the consistent and covariant anomalies
and compare them.

%%%%%%%%%%%%%%%%%%%%%%%%%%%%%%%%%%%%%%%%%%%%%%%%%%%%%%%%%%%%%%%%%%%%%
%%%%%%%%%%%%%%%%%%%%%%%%%%%%%%   SEC  2    %%%%%%%%%%%%%%%%%%%%%%%%%%
%%%%%%%%%%%%%%%%%%%%%%%%%%%%%%%%%%%%%%%%%%%%%%%%%%%%%%%%%%%%%%%%%%%%%
\section{Gauge symmetry of the heat-kernel }

%%%%%%%%%%%%%%%%%%%%%%%%%%%%%%%%%%%%%%%%%%%%%%%%%%%%%%%%%%%%%%%%%%%%%
%%%%%%%%%%%%%%%%%%%%%%%%%%%%%%%%%%%%%%%%%%%%%%%%%%%%%%%%%%%%%%%%%%%%%
%%%%%%%%%%%%%%%%%%%%%%%%%%%%%%%%%%%%%%%%%%%%%%%%%%%%%%%%%%%%%%%%%%%%%
\def\deri{{\cal D}}
%%%%%%%%%%%%%%%%%%%%%%%%%%%%%%%%%%%%%%%%%%%%%%%%%%%%%%%%%%%%%%%%%%%%%
%%%%%%%%%%%%%%%%%%%%%%%%%%%%%%%%%%%%%%%%%%%%%%%%%%%%%%%%%%%%%%%%%%%%%
%%%%%%%%%%%%%%%%%%%%%%%%%%%%%%%%%%%%%%%%%%%%%%%%%%%%%%%%%%%%%%%%%%%%%

We consider the bosonic and fermionic cases separately.
Focussing initially on the bosonic case.
The 1-loop action in n-dim space is generally represented as
%%%%%%%%%%%%   Baction %%%%%%%%%%%%%%%% 
\begin{eqnarray}
S[\p,\deri_b] 
= \int d^n x \frac{1}{2} \phi^{\dag} \deri_b \phi,
\label{Baction}
\end{eqnarray}
where $\phi$ is a bosonic quantum field and $\deri_b$ is written 
in terms of background fields.
(Here we use the matrix notation:\ 
$\phi^{\dag}\deri_b\phi=\phi^{\dag}_i\deri^{ij}_b\phi_j$
where $i,j$ are field-suffixes.)
Let $S[\p,\deri_b]$ be  invariant for a general local
transformation.
%** GS.2%%%%%%%%%%%%%%%%%%%%%%%%%%%%%%%%%%%%%%%%
\begin{eqnarray}
\phi^{\prime}  &=&  e^{-\Sigma(x)} \phi\com\q \nn \\
\phi^{\dag\prime}  &=&  \phi^{\dag} e^{-\Sigma(x)^{\dag}}\com\q\nn\\
\deri_b^{\prime}  &=&  e^{\Sigma(x)^{\dag}} \deri_b e^{\Sigma(x)}\com
\label{GS.2}
\end{eqnarray}
where $\Sigma(x)$ is a  local parameter
(local gauge, general coordinate transformation, local Weyl).
(In the case of the general coordinate transformation, $\Sigma(x)$
has derivatives).
We analyze the anomaly for the above general local symmetry.
The partition function,
%%%%%%%%%%%%%%%%%%%%%%%%%%%%%%%%%%%%%%%
\begin{eqnarray}
Z=\int\Dcal\phi e^{-S[\p,\deri_b]}\com
\end{eqnarray}
generally changes under the above symmetry as
%****(f.4)%%%%%%%%%%%%%%%%%%%%
\begin{eqnarray}
Z^{\prime} = \int\Dcal\phi^{\prime} 
e^{-S[\p^{\prime},\deri_b^{\prime}]}
= \int \Dcal\phi(x)~\det\frac{\pl\phi'(y)}{\pl\phi(x)}
e^{-S[\p,\deri_b]}\com
\label{f.4}
\end{eqnarray}
%%%%%%%%%%%%%%%%%%%%%%%%%%%%%
due to a change in the measure which is identified with the anomaly\cite{F79}.
The Jacobian is formally written as
%****(f.5)%%%%%%%%%%%%%%%%%%%%
\begin{eqnarray}
J\equiv
\det\frac{\pl\phi'(y)}{\pl\phi(x)}=\det~(e^{-\Sigma(x)}\del^n(x-y)~)
=\exp(-{\rm Tr}~[\Sigma(x)\del^n(x-y)]+O(\Sigma^2)),
\label{f.5}
\end{eqnarray}
%%%%%%%%%%%%%%%%%%%%%%%%%%%%%
where the infinitesimal $\Sigma(x)$ is taken in the last expression.
The anomaly is represented by (\ref{f.5}).

Next, we consider the fermion case. The 1-loop action in n-dim space
is written as 
%***   Faction  %%%%%%%%%%%%%%%%%%%%%%%%%%%%%%%%%%%%%%
\begin{eqnarray}
S[\psi,\barpsi;\Dcal_f] 
= \int d^n x\ i {\bar\psi} \deri_f \psi,
\label{Faction}
\end{eqnarray}
where $\psi$ is a (Dirac) fermionic quantum field and
${\bar\psi} = \psi^{\dag} \gamma^4$.
\footnote{
\ $\ga^4$ is  understood as the n-dim generalization of  $\ga^4$
\ in 4 dimension.
}
$\deri_f$ is written in terms of
background fields and is generally not hermitian.
Let this action be invariant for a local transformation.
%** GS.7 %%%%%%%%%%%%%%%%      %%%%%%%%%%%%%%%%%%%%%%%%%%%
\begin{eqnarray}
\psi^{\prime}  =  e^{-\Sigma(x)} \psi\com\q
{\bar\psi}^{\prime}  =  {\bar\psi} \gamma^4 
e^{-\Sigma(x)^{\dag}} \gamma^4\com\q
\deri_f^{\prime}  =  \gamma^4 e^{ \Sigma(x)^{\dag}} \gamma^4 
\deri_f e^{\Sigma(x)}.
\label{GS.7}
\end{eqnarray}
The anomaly for the above symmetry is given, as in the bosonic case, 
by the following Jacobian of the above local transformation:
%***** ferjaco %%%%%%%%%%%%%%%%%%%%%
\begin{eqnarray}
J &\equiv&
\left[ \det\frac{(\partial\psi'(y), \partial{\barpsi'}(y))}
{(\partial\psi(x), \partial{\barpsi}(x))}
\right]^{-1}
=\det~(e^{\Sigma(x)}\del^n(x-y)~)\det~(\gamma^4 e^{\Sigma^{\dag}(x)} \gamma^4 
{\del^n(x-y)}' ~)\nn\\
&=& \exp({\rm Tr}~[
\Sigma(x)\del^n(x-y)+ \gamma^4 \Sigma^{\dag}(x) \gamma^4 {\del^n(x-y)}'
]+O(\Sigma^2))\com
\label{ferjaco}
\end{eqnarray}
where ${\del^n(x-y)}'$ is used to note that it can be
different from $\del^n(x-y)$ at the regularization level.

In order to regularize the delta function $\del^n(x-y)$ appearing
both in (\ref{f.5}) and in (\ref{ferjaco}),
we introduce the heat-kernel,
%****(f.6)%%%%%%%%%%%%%%%%%%%%
\begin{eqnarray}
&&G(x,y;t)\equiv <x|e^{-t\Dvec}|y>,\q t>0,
\end{eqnarray}
using a properly chosen  elliptic differential operator
$\Dvec$.
The heat-kernel $G(x,y;t)$ , introduced ``symbolically'' above, 
is more precisely
defined by
%****heatequation%%%%%%%%%%%%%%%%%%%%
\begin{eqnarray}
&&(\frac{\pl}{\pl t}+\Dvec_x)G(x,y;t)=0,\q 
G(x,y;t)({\overleftarrow {\frac{\pl}{\pl t}}}
+{\overleftarrow D}^{\dag_y})=0,\q
t>0, \nn \\
&&\lim_{t\ra +0}~G(x,y;t)=\delta^n(x-y),
\label{heatequation}
\end{eqnarray}
where $G(x,y;t)$ is defined symmetrically with respect to $x$ and
$y$.
The last equation of (\ref{heatequation}) guarantees
that we can use $G(x,y;t)$ to regularize
$\del^n(x-y)$. 
Then the anomaly is obtained as \cite{II},
%**GS.11,12 %%%%%%%%%%%%%%%%%%%%%%%%%%%%%%%%%%%%%%%
\begin{eqnarray}
\ln J = - \lim_{t \rightarrow +0} {\rm Tr} [ \Sigma(x) G(x,y;t)], 
\q \mbox{for the bosonic case},\label{GS.11}\\
\ln J = \lim_{t \rightarrow +0} {\rm Tr}~[
\Sigma(x)G(x,y;t)+ \gamma^4\Sigma^{\dag}(x) \gamma^4 {G(x,y;t)}'
],
\q \mbox{for the fermionic case}.\label{GS.12}
\end{eqnarray}
where ${G(x,y;t)}'\equiv <x|e^{-t\Dvec'}|y>$ is generally a
heat-kernel with some
another operator $\Dvec'$.

In order for the heat kernel to be well-defined, 
and to be defined by the
action of (\ref{Baction}) or (\ref{Faction})
we require the following conditions for $\Dvec$:\ 
%%%1)\ $\Dvec$ is hermitian;
$\Dvec$ is  a second order elliptic differetial operator;\ 
$\Dvec$ is made of the differential operators appearing
in the field equation. We do not impose hermiticity on $\Dvec$.
(See Sect.6 for further discussion).
For example, in the bosonic case, we can take 
%%%%%%%%%%%%%%%%%%%%%%%%%%%%%%%%%%%%%%%%
\begin{eqnarray}
\Dvec = \deri_b=\deri_b^{\dag}\pr
\end{eqnarray}
In the fermionic case, we have two cases.
%%%%%%%%%%%%%%%%%%%%%%%%%%%%%%%%%%%%%%%%%
\begin{eqnarray}
& \mbox{$\deri_f$ is hermitian}\q\q &
                              \q\q \Dvec =\deri_f \deri_f  \nn\\
& \mbox{$\deri_f$ is not hermitian}\q\q &
\Dvec = \left\{
\begin{array}{ll}
\deri_f \deri_f\ \ ,\ \ \deri_f^{\dag}\deri_f^{\dag}\ \ , & \\
\deri_f^{\dag} \deri_f\ \ ,\ \ \deri_f\deri_f^{\dag}\ \ , & \\
\half(\deri_f^{\dag} \deri_f +\deri_f \deri_f^{\dag})\ \ , &\mbox{etc.} 
\end{array}
\right.
\end{eqnarray}
In the non-hermitian case,
the operators
$\deri_f^2\ \mbox{and}\ (\deri_f^{\dag})^2$\  
are not hermitian.  
The different choices of the heat kernel operator correspond to 
the different choices of
reguralization.
It causes the different anomalies, in 
appearance, which are examined in a later section.

We denote the basis of the Hilbert space, on which
$\deri_i$ ($i=b\mbox{ or }f$)
operates, as  $\{f_n (x)\}$. 
The Hermite conjugate of $\deri_i$, appearing above, is
defined precisely as follows:
%*** prod %%%%%%%%%%%%%%%%%%%%%%%%%%%%%%%%%%%%%%%%%%
\begin{eqnarray}
(f_m, \deri_i f_n )= (\deri^{\dag}_i f_m, f_n),
\label{prod}
\end{eqnarray}
where $(\Phi,\Psi)\equiv \int d^nx\Phi^{\dag}(x)\Psi(x)$ 
is an inner product on the Hilbert space.
On this space, let us introduce the following local gauge
transformation:
%*** gengauge  %%%%%%%%%%%%%%%%%%%%%%%%%%%%%%%%%%%%%%
\begin{eqnarray}
%\phi^{\prime} & = & e^{-\Lambda(x)} \phi, \nn \\ 
%\phi^{\dag\prime} & = & \phi^{\dag} e^{\Lambda(x)} , \nn \\
f_n^{\prime}(x) =  e^{-\Lambda(x)} f_n (x)\com\q
f_n^{\dag\prime}(x)  =  f_n^{\dag}(x) e^{\Lambda(x)}\com\q
\deri_i^{\prime}  =  e^{-\Lambda(x)} \deri_i e^{\Lambda(x)}\com
\label{gengauge}
\end{eqnarray}
where $\La(x)$ is the anti-hermitian gauge parameter:\ 
%%%%%%%%%%%%%%%%%%%%%%%%%%%%%%%%%%%%%%%%
%\begin{equation}
$
\Lambda(x)^{\dag} = - \Lambda(x)
$.
%\end{equation}
For the above transformation (\ref{gengauge}), 
the matrix element of $\deri_i$, defined by the inner product
(\ref{prod}), does not change. 

In both boson and fermion cases, 
$\Dvec$ transforms as
%%%%%%%%%%%%%%%%%%%%%%%%%%%%%%%%%%%%%%%%%%%%%%%%%
\begin{eqnarray}
\Dvec_x^{\prime} & = & e^{-\Lambda(x)} \Dvec_x e^{\Lambda(x)}\com\q
{\overleftarrow D}^{\dag\prime}_x = e^{-\Lambda(x)} 
{\overleftarrow D}^{\dag}_x e^{\Lambda(x)}.
\label{transd}
\end{eqnarray}
Because the heat kernel is defined by the operator $\Dvec$
, through (\ref{heatequation}), and $\Dvec$ changes as above, 
the heat kernel must itself change.
The heat kernel equation transforms as follows:
\begin{eqnarray}
(\frac{\pl}{\pl t}+\Dvec_x^{\prime})G^{\prime}(x,y;t)=0, \q
G^{\prime}(x,y;t)({\overleftarrow {\frac{\pl}{\pl t}}}
        +{\overleftarrow D}^{\dag\prime}_y)=0,             
\label{transg1}
\end{eqnarray}
which implies
%**** transg2 %%%%%%%%%%%%%%%%%%%%%
\begin{eqnarray}
e^{-\Lambda(x)} (\frac{\pl}{\pl t}+\Dvec_x) e^{\Lambda(x)} 
G^{\prime} (x,y;t) =0, \q
G^{\prime} (x,y;t) e^{-\Lambda(y)} ({\overleftarrow {\frac{\pl}{\pl t}}}
+{\overleftarrow D}^{\dag}_y) e^{\Lambda(y)} = 0,
\label{transg2}
\end{eqnarray}
where $G^{\prime}(x,y;t)$ is the transform of $G(x,y;t)$.
Therefore, by (\ref{heatequation}) and (\ref{transg2}), the solution
$G(x, y; t)$ has to transform covariantly as follows:
%****  transg  %%%%%%%%%%%%%%%%%%%
\begin{eqnarray}
G^{\prime}(x, y; t) & = & e^{-\Lambda(x)} G(x, y; t)  e^{\Lambda(y)}.
\label{transg}
\end{eqnarray}

%%%%%%%%%%%%%%%%%%%%%%%%%%%%%%%%%%%%%%%%%%%%%%%%%%%%%%%%%%%%%%%%%%%%%
%%%%%%%%%%%%%%%%%%%%%%%%%%%%%%   SEC  3    %%%%%%%%%%%%%%%%%%%%%%%%%%
%%%%%%%%%%%%%%%%%%%%%%%%%%%%%%%%%%%%%%%%%%%%%%%%%%%%%%%%%%%%%%%%%%%%%
\section{Covariants and Covariant Derivative}
We now move on to the gauge symmetry of the anomaly formulae.
The formulae are obtained by calculating $x=y$ part (trace part). 
So we consider the case of
$x=y$ in (\ref{transg}). If we take the infinitesimal form,
(\ref{transd}) and (\ref{transg}) become
%***inftransd %%%%%%%%%%%%%%%%%%%%%%%%%%%%%%%%%%%%%%
\begin{eqnarray}
\gsym \Dvec_x = - [ \Lambda(x), \Dvec_x ]\com\q 
\gsym G(x, x; t)  =  - [ \Lambda(x), G(x, x; t) ]\pr
\label{inftransd}
\end{eqnarray}
Generally, $\Dvec_x$ is a second order differential operator, so 
its most general form is written as 
%**cov.3 %%%%%%%%%%%%%%%%%%%%%%%%%%%%%%%%%%%%
\begin{eqnarray}
\Dvec_x = - (\delta_{\mu\nu} + W_{\mu\nu}(x) ) \partial_\mu \partial_\nu 
- N_\mu(x) \partial_\mu - M(x),\label{cov.3}
\end{eqnarray}
where $W_{\mu\nu}(x)=W_{\nu\mu}(x)$, $N_\mu(x)$ and $M(x)$ 
are represented by background fields.\footnote{
$\del_\mn +W_\mn(x)$ is the metric of the present general system.
}
(
The above equation is written, in component form, as
$
\Dvec_x^{ij} = - (\delta_{\mu\nu}\del^{ij}
                  + W_{\mu\nu}^{ij}(x) ) \partial_\mu \partial_\nu 
- N_\mu^{ij}(x) \partial_\mu - M^{ij}(x),
$
where $i$ and $j$ are all suffixes of 
fields; $i,j=1,2,\cdots, N$.)
~From (\ref{inftransd}), we can derive the transformation of
$W_{\mu\nu}(x)$, $N_\mu(x)$ and $M(x)$ as
%*trawnm%%%%%%%%%%%%%%%%%%%%%%%%%%%%%%%%%%%
\begin{eqnarray}
\gsym W_{\mu\nu} & = & - [ \Lambda, W_{\mu\nu} ], \nn \\
\gsym N_{\mu} & = & 2 (\delta_{\mu\nu} + W_{\mu\nu} ) \partial_\nu \Lambda 
- [ \Lambda, N_\mu ], \nn \\
\gsym M & = & (\delta_{\mu\nu} + W_{\mu\nu} ) 
\partial_\mu \partial_\nu \Lambda 
+ N_\mu \partial_\mu \Lambda - [ \Lambda, M ].
\label{trawnm}
\end{eqnarray}

We assume that the 'background coefficient' 
of the second order differential terms
$(\delta_{\mu\nu} + W_{\mu\nu}(x) )$ is non-degenerate. It is
necessary for the propagator of $\phi(x)$ or $\psi(x)$\ to be defined.
We can define $Z_{\mu\nu}$ as 
%*zdef %%%%%%%%%%%%%%%%%%%%%%%%%%%%%%%%%%
\begin{equation}
(\delta_{\mu\nu} + W_{\mu\nu} )(\delta_{\nu\lambda} + Z_{\nu\lambda} )
= \delta_{\mu\lambda},
\label{zdef}
\end{equation}
where $Z_{\mu\nu} = Z_{\nu\mu}$.
It is expressed as an infinite series in $W_{\mu\nu}$ by
%*cov.6 %%%%%%%%%%%%%%%%%%%%%%%%%%%%%%%%%
\begin{eqnarray}
Z_{\mu\nu} = - W_{\mu\nu} + W_{\mu\lambda} W_{\lambda\nu} 
- W_{\mu\lambda} W_{\lambda\rho} W_{\rho\nu} + O(W^4).
\label{cov.6}
\end{eqnarray}
$Z_{\mu\nu}$ transforms covariantly, as derived from (\ref{zdef}),
%* cov.7 %%%%%%%%%%%%%%%%%%%%%%%%%%%%%%%%%%
\begin{eqnarray}
\gsym Z_{\mu\nu} = - [ \Lambda, Z_{\mu\nu} ].
\label{cov.7}
\end{eqnarray}
We can find the generalized gauge field $A_\mu$ 
of the present general gauge transformation as,
%** adef %%%%%%%%%%%%%%%%%%%%%%%%%%%%%%%%%%%
\begin{eqnarray}
A_\mu \equiv \half (\delta_{\mu\nu} + Z_{\mu\nu} ) N_\nu\com
\label{adef}
\end{eqnarray}
which surely transforms, from (\ref{trawnm}) and
(\ref{adef}), as a gauge field.
%* cov.9 %%%%%%%%%%%%%%%%%%%%%%%%%%%%%%%%%%
\begin{eqnarray}
\gsym A_\mu = \partial_\mu \Lambda - [ \Lambda, A_\mu ].
\label{cov.9}
\end{eqnarray}

We want to obtain the solution $G(x, x; t)$, expressed by
$W_{\mu\nu}(x)$, $N_\mu(x)$\ and $M(x)$. 
Since $G(x, x; t)$ is covariant with respect to the gauge
transformation (\ref{transg}) or (\ref{trawnm}), 
it must be constructed by covariant quantities.
We find them, by (\ref{trawnm}),(\ref{cov.7}) and (\ref{cov.9}), as
%%%%%%%%%%%%%%%%%%%%%%%%%%%%%%%%%
\begin{eqnarray}
X \equiv M - (\delta_{\mu\nu} + W_{\mu\nu} ) \partial_\mu A_\nu
- (\delta_{\mu\nu} + W_{\mu\nu} ) A_\mu A_\nu\ \ ,\ 
Y_{\mu\nu} \equiv \partial_\mu A_\nu 
- \partial_\nu A_\mu + [ A_\mu, A_\nu ]\ \ ,\ 
W_{\mu\nu}\ .
\end{eqnarray}
Indeed, $X$ and $Y_{\mu\nu}$ transform covariantly,
%%%%%%%%%%%%%%%%%%%%%%%%%%%%%%%%%%%
\begin{eqnarray}
\gsym X = - [ \Lambda, X]\com\q
\gsym Y_{\mu\nu} = - [ \Lambda, Y_{\mu\nu} ]\pr
\end{eqnarray}
Those covariants were examined in the counterterm formula,  
for the special case of $W_\mn=0$ by \cite{tH}
and for the general case by \cite{I92}.
Moreover we can define a covariant derivative $D_\mu$ 
on a covariant $\Phi$, as
%%%%%%%%%%%%%%%%%%%%%%%%%%%%%%%%%%
\begin{eqnarray}
D_\mu \Phi \equiv \partial_\mu \Phi + [ A_\mu, \Phi ].
\end{eqnarray}
Their transformations are
%%%%%%%%%%%%%%%%%%%%%%%%%%%%%%%%%
\begin{eqnarray}
\gsym \Phi = - [ \Lambda, \Phi]\com\q
\gsym (D_\mu \Phi) = - [ \Lambda, D_\mu \Phi]\pr
\end{eqnarray}
Therefore we can conclude that $G(x, x; t)$ 
must be written by the covariant quantities,
$X(x)$, $Y_{\mu\nu}(x)$, $W_{\mu\nu}(x)$ and $D_\mu(x)$.
%%%%%%%%%%%%%%%%%%%%%%%%%%%%%%%%%%%
\begin{eqnarray}
G(x, x; t) = F(X(x), Y_{\mu\nu}(x), W_{\mu\nu}(x), D_\mu(x)),
\end{eqnarray}
where $F(X, Y, W, D)$ is some function which will be determined
in the next section.
For later use, we note mass-dimension of the covariants:\ 
$[X]=[Y_\mn]=(\mbox{Mass})^2,[D_\m]=\mbox{Mass},[W_\mn]=(\mbox{Mass})^0$.
%%%%%%%%%%%%%%%%%%%%%%%%%%%%%%%%%%%%%%%%%%%%%%%%%%%%%%%%%%%%%%%%%%%%%
%%%%%%%%%%%%%%%%%%%%%%%%%%%%%%   SEC  4    %%%%%%%%%%%%%%%%%%%%%%%%%%
%%%%%%%%%%%%%%%%%%%%%%%%%%%%%%%%%%%%%%%%%%%%%%%%%%%%%%%%%%%%%%%%%%%%%
\section{Anomaly Formulae}
Decomposing the operator $\Dvec_x$ as\ 
%****(f.10)%%%%%%%%%%%%%%%%%%%%
%\begin{eqnarray}
$
\Dvec_x = -\del_\mn \pl_\m\pl_\n-\Vvec(x),\ 
\Vvec (x)  \equiv 
W_\mn (x)\pl_\m\pl_\n+N_\m \pl_\m+M$, 
%\label{f.10}
%\end{eqnarray}
%%%%%%%%%%%%%%%%%%%%%%%%%%%%%%
the heat kernel equation (\ref{heatequation}) can be written as 
%****(f.11)%%%%%%%%%%%%%%%%%%%%
\begin{eqnarray}
(\frac{\pl}{\pl t}-\pl^2)G (x,y;t) =
\Vvec (x)G (x,y;t),
%\pl^2\equiv \del_\mn\pl_\m\pl_\n=\sum^n_{\mu=1}(\frac{\pl}{\pl
%x^\m})^2.
                                                          \label{f.11}
\end{eqnarray}
We solve this equation by the propagator approach\cite{BD,II}.
The solution of
%\begin{flushleft}
%i)\ Heat Equation
%\end{flushleft}
the  heat equation is given by
%****(f.12)%%%%%%%%%%%%%%%%%%%%
\begin{eqnarray}
& (\frac{\pl}{\pl t}-\pl^2)G_0(x,y;t)=0\com\q 
G_0(x,y;t)
=\frac{1}{(4\pi t)^{n/2}}e^{-\frac{(x-y)^2}{4t}}~I_N\com\q
                                         \mbox{for}\  t>0\com\q & \nn\\
& G_0(x,y;t)\equiv 0 \com\q \mbox{ for } t\leq 0\com\q & 
\label{f.12}
\end{eqnarray}
%%%%%%%%%%%%%%%%%%%%%%%%%%%%%
%has the solution
%****(f.13)%%%%%%%%%%%%%%%%%%%%
%\begin{eqnarray}
%G_0(x,y;t)=G_0(x-y;t)=\int\frac{d^nk}{(2\pi)^n}\exp\{-k^2t+ik^\m(x-y)^\m\}
%                                              ~I_N\nn\\
%=\frac{1}{(4\pi t)^{n/2}}e^{-\frac{(x-y)^2}{4t}}~I_N\com\q
%k^2\equiv\sum^n_{\m=1}(k^\m)^2\com \label{f.13}
%\end{eqnarray}
%%%%%%%%%%%%%%%%%%%%%%%%%%%%%
where $I_N$ is the $N\times N$ identity matrix and
$G_0$ satisfies the initial condition:\ $
\lim_{t\ra +0}~G_0(x-y;t)=\del^n(x-y)~I_N
$\ .
%We define
%****(f.13b)%%%%%%%%%%%%%%%%%%%%
%\begin{eqnarray}
%G_0(x,y;t)=0 \mbox{ for } t\leq 0.\label{f.13b}
%\end{eqnarray}
%%%%%%%%%%%%%%%%%%%%%%%%%%%%%
%%%%%%%%%%%%%%%%%%%%%%%%  ii)  %%%%%%%%%%%%%%%%%%%%%%%%%%%%%%%%%%%%%%
%\begin{flushleft}
%ii)\ Heat Propagator
%\end{flushleft}
The heat equation with the delta-function source defines the heat
propagator and which is given by.
%****(f.14)%%%%%%%%%%%%%%%%%%%%
\begin{eqnarray}
(\frac{\pl}{\pl t}-\pl^2)S(x,y;t-s)=\del(t-s)\del^n(x-y)~I_N\com\nn\\
S(x,y;t)
%=S(x-y;t)=\int\frac{d^nk}{(2\pi)^n}\frac{dk^0}{2\pi}
%\frac{\exp\{-ik^0t+ik\cdot(x-y)\} }{-ik^0+k^2}~I_N  \nn\\
=\th(t)G_0(x-y;t)\com
                                                           \label{f.14}
%k^2\equiv \sum^\m_{\n=1}k^\m k^\m\com\q
%k\cdot x\equiv \sum^n_{\m=1}k^\m x^\m\pr\nn
\end{eqnarray}
%%%%%%%%%%%%%%%%%%%%%%%%%%%%%
where $\th(t)$\ is the {\it step function} defined by
$\th(t)=1\ \mbox{for}\ t>0\ ,\ \th(t)=0\ \mbox{for}\ t<0\ $.
%$S(x-y;t)$\ satisfies the initial condition:\ $
%\lim_{t\ra+0}S(x-y;t)=\del^n(x-y)~I_N$\
%and $
%S(x,y;t)=0 \mbox{ for } t\leq 0$\ .
Using the above quantities, the formal solution of (\ref{f.11})
(or (\ref{heatequation}))
%with the initial condition (\ref{}) 
is given by
%****(f.15)%%%%%%%%%%%%%%%%%%%%
\begin{eqnarray}
G(x,y;t)=G_0(x-y;t)+\int d^nz\int^\infty_{-\infty}ds~
S(x-z;t-s)\Vvec(z)G(z,y;s)               \pr\label{f.15}
\end{eqnarray}
The iterative solution of (\ref{f.15}) 
with respect to the order of $\Vvec(x)$
is obtained in \cite{II}.
%The solution is iteratively obtained as 
%****(f.16)%%%%%%%%%%%%%%%%%%%%
%\begin{eqnarray}
%G(x,y;t)&=&
%G_0(x-y;t)+\int S\Vvec G_0+\int S\Vvec\int S\Vvec G_0+\cdots\com\nn\\
%G_1(x,y;t)&\equiv&\int S\Vvec G_0=\int d^nzds
%S(x-z;t-s)\Vvec(z)G_0(z-y;s)\nn\\
%&=&\int d^nz\int^t_0ds
%G_0(x-z;t-s)\Vvec(z)G_0(z-y;s)\com\label{f.16}\\
%G_2(x,y;t)&\equiv&\int S\Vvec\int S\Vvec G_0
%=\int d^nz'ds'S(x-z';t-s')\Vvec(z')\nn\\
%&&\times\int d^nz ds S(z'-z;s'-s)\Vvec(z)G_0(z-y;s)\nn\\
%&=&\int d^nz'\int^t_0ds' G_0(x-z';t-s')\Vvec(z')\nn\\
%&&\times\int d^nz\int^{s'}_0ds G_0(z'-z;s'-s)\Vvec(z)G_0(z-y;s)\pr\nn
%\end{eqnarray}
%%%%%%%%%%%%%%%%%%%%%%%%%%%%%
%Higher-order terms are similarly obtained.
The trace-part of the solution $G(x, y; t)$
%(\ref{f.5})
is calculated from the diagonal-part of the solution.
%****(f.17)%%%%%%%%%%%%%%%%%%%%
\begin{eqnarray}
G(x,x;t) &=& G_0(0;t)+G_1(x,x;t)+G_2(x,x;t)+\cdots,
\end{eqnarray}
where the suffix numbers show the iteration order.
Generally, in n dimensions, the terms up to $G_{n/2}$\ are practically
sufficient for the anomaly calculation.

%G_0(0;t) &=&\frac{1}{(4\pi t)^{n/2}}~I_N.
%G_1(x,x;t)&\equiv&\int \left.S\Vvec G_0\right|_{x=y}
%\label{f.17}\\
%&=&\frac{1}{(4\pi)^nt^{(n/2)-1}}\int
%d^nw\int^1_0dr\frac{1}{\{(1-r)r\}^{n/2}}
%e^{-\frac{w^2}{4(1-r)}}~\Vvec(x+\sqrt{t}w)e^{-\frac{w^2}{4r}},
%                                                 \nn\\
%\Vvec(x+\sqrt{t}w)&=&
%\frac{1}{t}W_\mn(x+\sqrt{t}w)\frac{\pl}{\pl w^\m}\frac{\pl}{\pl w^\n}
%+\frac{1}{\sqrt{t}}N_\m(x+\sqrt{t}w)\frac{\pl}{\pl
%w^\m}+M(x+\sqrt{t}w)\ ,\nn
%\end{eqnarray}
%%%%%%%%%%%%%%%%%%%%%%%%%%%%%%
%where we introduce some scaled integration variables which are
%dimension-less:\\
%
%$r=s/t, w^\mu=(z-x)^\mu/\sqrt{t}\ $. Furthermore
%%****(f.18)%%%%%%%%%%%%%%%%%%%%
%\begin{eqnarray}
%&&G_2(x,x;t)\equiv\left.\int S\Vvec\int S\Vvec G_0\right|_{x=y}\nn\\
%&=&\frac{1}{(4\pi)^{3n/2}}\frac{1}{t^{(n/2)-2}}\int d^nv \int d^nu
%\int^1_0dk\int^k_0dl\frac{1}{\{(1-k)(k-l)l\}^{n/2}}
%e^{-\frac{v^2}{4(1-k)}}                    \label{f.18}\\
%&&\times \Vvec(x+\sqrt{t}v)e^{-\frac{(v-u)^2}{4(k-l)}}
%\Vvec(x+\sqrt{t}u)e^{-\frac{u^2}{4l}}\com\nn
%\end{eqnarray}
%%%%%%%%%%%%%%%%%%%%%%%%%%%%%
%where $k=s'/t\ ,\ l=s/t\ ,\ v^\m=(z'-x)^\m/\sqrt{t}\ ,\ u^\m=
%(z-x)^\m/\sqrt{t}$.
Further analysis will be done for each dimension.
For simplicity, we assume that $W_{\mu\nu}$ commutes with
$W_{\mu\nu}$, $N_\mu$ and $M$. This assumption is
valid for most field theories in flat and curved spacetime.\footnote{
This is the case $W_\mn^{ij}\propto \del^{ij}$.
The case that $W$ does not commute with $W,N,M$ takes place, for example,
when we consider Weyl anomaly in
the  gravity-photon coupling system with a non-Feynman-like 
gage:
$\Lcal_{int}=\half\sqg \{ A_\la (g^\ls \na^2-\al \na^\la\na^\si)A_\si\},
\al\neq 0$.
         }
Only the term of $G(x, x; t) |_{t^0}$ 
is necessary for the anomaly calculation,
because we take the limit $t\ra +0$.
Divergent parts of $O(t^{-m})$\ ($m=1,2,\cdots >0$)\ are considered
to be renormalized. 
By using the gauge symmetry found in the previous section, we can
express the anomaly formulas obtained in \cite{II} in a
covariant way. Let us consider the 2-dim and 4-dim cases.

\begin{flushleft}
(i) Two Dimensions (n=2)
\end{flushleft}
First for the special case of $W_\mn=0$,
%**f.47%%%%%%%%%%%%%%%%%%%%%%%%%%%%%%%%%%
\begin{eqnarray}
G(x, x; t) |_{t^0} =
\frac{1}{4 \pi} X \com\q
W_\mn=0\com \q X=M-\half \pl_\m N_\m-\frac{1}{4}N_\m N_\m\pr 
\label{f.47}
\end{eqnarray}
For this case,  $G(x, x; t) |_{t^0}$ is completely fixed.
For the general case of $W_\mn$, we obtain 
%**f.46 %%%%%%%%%%%%%%%%%%%%%%%%%%%%%%%%%
\begin{eqnarray}
G(x, x; t) |_{t^0} &=&
\frac{1}{4 \pi} \left[ X - \frac{1}{12} D^2 W_{\mu\mu} 
+ \frac{1}{3} D_\mu D_\nu W_{\mu\nu} + 
O((W,X,Y)^2) \right],
\label{f.46}
\end{eqnarray}
where $O((W,X,Y)^3)$ are, from the dimensional reason, 
restricted to the following forms. 
%**f.46b %%%%%%%%%%%%%%%%%%%%%%%%%%%%%%%%%
\begin{eqnarray}
X\cdot O_1(W)\com\q Y\cdot O_2(W)\com\q DDW\cdot O_3(W)\com\q DW\cdot
DW\cdot O_4(W^0)
\label{f.46b}
\end{eqnarray}
where $O_i(W^s)$ ($i=1\sim 4; s=0,1$) are some functions made only of $W$ with
respective orders of $W^s$.

\begin{flushleft}
(ii) Four Dimensions (n=4)
\end{flushleft}
In the four dimension, we obtain the following formulas. 
For  the case of $W_\mn=0$, we obtain a complete expression.
%**** 4dformW0  %%%%%%%%%%%%%%%%%%%%%%%
\begin{eqnarray}
G(x, x; t) |_{t^0} =\frac{1}{(4 \pi)^2} 
\biggl[ \frac{1}{6} D^2 X 
+ \frac{1}{2} X^2 + \frac{1}{12} Y_{\mu\nu} Y_{\mu\nu} \biggr]\com\q
D_\m X=\pl_\m X+\half [N_\m,X]\com                  \nn\\
X=M-\half\pl_\m N_\m-\frac{1}{4}N_\m N_\m\com\q
Y_\mn=\half(\pl_\m N_\n-\pl_\n N_\m)+\frac{1}{4}[N_\m,N_\n]\pr
\label{4dformW0}
\end{eqnarray}
The first total derivative term is the difference from
the 1-loop counter-term formula\cite{tH}. This formula will be used
in the next section.
For the general case of $W_\mn$, we have
%** 4dformula %%%%%%%%%%%%  4dformula  %%%%%%%%%%%%%%%%%%%%%%%
\begin{eqnarray}
G(x, x; t) |_{t^0} &=&
\frac{1}{(4 \pi)^2} 
\biggl[ \frac{1}{6} D^2 X - \frac{1}{120} D^2 D^2 W_{\mu\mu} 
+ \frac{1}{20} D^2 D_\mu D_\nu W_{\mu\nu} \nn \\
&&+ \frac{1}{2} X^2 + \frac{1}{12} Y_{\mu\nu} Y_{\mu\nu} 
 - \frac{1}{12} D^2 W_{\mu\mu}\cdot X 
+ \frac{1}{3} D_\mu D_\nu W_{\mu\nu}\cdot X 
- \frac{1}{6} D_\lambda D_\mu W_{\lambda\nu} \cdot Y_{\mu\nu}  \nn \\
&& - \frac{1}{18} D_\lambda W_{\mu\mu}\cdot D_\lambda X
- \frac{1}{12} W_{\mu\mu} D^2 X
+ \frac{1}{6} D_\mu W_{\mu\nu}\cdot D_\nu X
+ \frac{1}{6} W_{\mu\nu} D_\mu D_\nu X \nn \\
&& + \frac{1}{6} D_\lambda W_{\lambda\nu}\cdot D_\mu Y_{\mu\nu}
+ 0 \times D_\mu W_{\lambda\nu}\cdot D_\lambda Y_{\mu\nu}
+ 0 \times W_{\lambda\nu} D_\lambda D_\mu Y_{\mu\nu} \nn \\
&&+(DDW\cdot DDW\mbox{-terms}, \mbox{see Table 1})
+(DW\cdot DDDW\mbox{-terms})     \nn \\
&&+(W\cdot DDDDW\mbox{-terms})+O((W,X,Y)^3)
  \biggr].
\label{4dformula}
\end{eqnarray}
Suppressed parts above are not necessary for the practical use.
Terms of $O((W,X,Y)^3)$ are restricted to some types, 
which can be determined as in n=2 case.
Graph names, used in Table 1, are defined in \cite{II}
where each expression is shown graphically.

\vspace{10pt}
\begin{tabular}{|c|c|c||c|c|c|}
\hline
Graph      & Expression  & Coeff
& Graph & Expression & Coeff   \\
\hline
$\bar {A1}$ & $D_\si D_\la W_\mn\cdot D_\si D_\n W_{\m\la}$ & 1/45 &
${\bar Q}^2$ & $(D_\m D_\n W_{\mn})^2$                        & 1/18\\
$\bar {A2}$ & $D_\si D_\la W_{\la\m}\cdot D_\si D_\n W_{\mn}$ & -2/45&
$\bar {C1}$ & $D_\m D_\n W_{\la\la}\cdot D_\m D_\n W_{\si\si}$ & 1/360\\
$\bar {A3}$ & $D_\si D_\la W_{\la\m}\cdot D_\m D_\n W_{\n\si}$ & -2/45&
$\bar {C2}$ & $D^2 W_{\mn}\cdot D^2 W_{\mn}$                   & 1/144 \\
$\bar {B1}$ & $D_\n D_\la W_{\si\si}\cdot D_\la D_\m W_{\mn}$ & -1/90&
$\bar {C3}$ & $D_\m D_\n W_{\la\la}\cdot D^2 W_{\mn}$         & -1/90 \\
$\bar {B2}$ & $D^2 W_{\la\n}\cdot D_\la D_\m W_{\mn}$        & 1/180&
${\bar P}{\bar Q}$ & $D^2 W_{\la\la}\cdot D_\m D_\n W_{\mn}$  & -1/36 \\
$\bar {B3}$ & $D_\m D_\n W_{\la\si}\cdot D_\m D_\n W_{\ls}$ & 1/180&
${\bar P}^2$ & $(D^2 W_{\la\la})^2$                             & 1/288 \\
$\bar {B4}$ & $D_\m D_\n W_{\la\si}\cdot D_\la D_\si W_{\mn}$ & 1/180&
                                                                   && \\
\hline
\multicolumn{6}{c}{\q}                                                 \\
\multicolumn{6}{c}{Table 1\ \  Anomaly Formula for $(DD W)^2$-part
                               of $G_2(x,x;t)|_{t^0}$ }\\
\multicolumn{6}{c}{\q\q\q The overall factor is $1/(4\pi)^2$.
Graph names are defined in \cite{II}    }\\
\end{tabular}
%%%%%%%%%%%%%%%%%%%%%%%%%  END  of  Table 1

When we consider an anomaly in a gravitational model, 
the general case of $W_\mn$ must be taken.  
Even in a flat theory,
the case appears when we evaluate a gauge-loop effect to
the (Weyl) anomaly in a gauge different from Feynman gauge.

%%%%%%%%%%%%%%%%%%%%%%%%%%%%%%%%%%%%%%%%%%%%%%%%%%%%%%%%%%%%%%%%%%%%%
%%%%%%%%%%%%%%%%%%%%%%%%%%%%%%   SEC  5    %%%%%%%%%%%%%%%%%%%%%%%%%%
%%%%%%%%%%%%%%%%%%%%%%%%%%%%%%%%%%%%%%%%%%%%%%%%%%%%%%%%%%%%%%%%%%%%%
\section{The Non-abelian Anomaly in Four Dimensions}

As a typical example, we consider the non-abelian anomaly of the
Dirac fields coupled to the V-A current in 4 dim space
\cite{B69,BIM,GJ,BZ}.
The Lagrangian is written as
%*** laglan %%%%%%%%%%%%%%%%%%%%%%%%%%
\begin{eqnarray}
{\cal L} & =& - {\bar\psi} \gamma^\mu ( \partial_\mu -i V_\mu 
-i A_\mu \gamma_5) \psi,
\label{laglan}
\end{eqnarray}
where $V_\mu\equiv T^aV_\m^a$ and $A_\mu\equiv T^aA_\m^a$ are gauge
fields and the convention:
$(\ga^\m)^{\dag}=\ga^\m,\ \ga_5^{\dag}=\ga_5$ ,
is taken.
We take the left and right decomposition:\ 
%%%%%%%%%%%%%%%%%%%%%%%%%%%%%%%%%%%%%
%\begin{eqnarray}
$
R_\mu = V_\mu + A_\mu,\ 
L_\mu = V_\mu - A_\mu
$.
%\end{eqnarray}
%
%
For simplicity, we set $L_\mu = 0$ and analyze only the $R_\mu$ part.
The Lagrangian is then represented as
%** laglan2 %%%%%%%%%%%%%%%%%%%%%%%%%%%%%%%%%%%%%
\begin{eqnarray}
{\cal L}_R = - {\bar\psi} \gamma^\mu ( \partial_\mu 
- i\frac{1+\gamma_5}{2} R_\mu) \psi
= - {\bar\psi} \Dslash_{R+} \psi,\label{laglan2}
\end{eqnarray}
where we define the following differential operators,
%%%%%%%%%%%%%%%%%%%%%%%%%%%%%%%%%%%%%
\begin{eqnarray}
\Dslash_R = \gamma^\mu (\partial_\mu - iR_\mu)\ ,\ \
\Dslash_{R+} 
= \gamma^\mu (\partial_\mu - i\frac{1+\gamma_5}{2} R_\mu)\ ,\ \ 
\Dslash_{R-} 
= \gamma^\mu (\partial_\mu - i\frac{1-\gamma_5}{2} R_\mu)
\ \ .
\end{eqnarray}
( We consider the case:\ 
$ \deri_f=i\Dslash_{R+} $
in (\ref{Faction})).
We note here $i\Dslash_R$ is hermitian, whereas $i\Dslash_{R+}$
is not\ $(i\Dslash_{R+})^{\dag}=i\Dslash_{R-}$.
The Lagrangian (\ref{laglan2}) is invariant under the following local
gauge symmetry, 
% ** rmugauge %%%%%%%%%%%%%%%%%%%%%%%%%%%%%%%%%%%%%%%
\begin{eqnarray}
\psi \longrightarrow \psi^\prime = e^{-i \lambda(x)
\frac{1+\gamma_5}{2} } \psi\com\q
{\bar\psi} \longrightarrow {\bar\psi}^\prime = {\bar\psi} e^{i \lambda(x)
\frac{1-\gamma_5}{2} }\com\nn \\
R_\m \longrightarrow R_\m^\prime=U(\la)^{-1}R_\m U(\la)
-i\pl_\m U(\la)^{-1}\cdot U(\la)\com\q U(\la)=e^{i\la(x)}\com
\label{rmugauge}
\end{eqnarray}
where $\lambda(x)\equiv T^a\la^a(x)$ is a gauge parameter.
We calculate the anomaly under the above symmetry.
%The Lagrangian has the above gauge symmetry.
~From (\ref{ferjaco}), the Jacobian of this transformation is written
as ($\Sigma=i\la(x)\frac{1+\ga_5}{2}$)
%%%%%%%%%%%%%%%%%%%%%%%%%%%%%%%%%%%%%%
\begin{eqnarray}
\ln J 
%= [\ln J]_{\psi} + [\ln J]_{\bar\psi} \nn \\
&=& i {\rm Tr} [ 
\lambda(x)(\frac{1+ \gamma_5}{2} \delta^4(x-y)
          -\frac{1-\ga_5}{2} {\del^4(x-y)}') 
             ]                          \nn \\
&=& i\lim_{t\ra +0} {\rm Tr} [ 
\lambda(x)(\frac{1+ \gamma_5}{2} <x|e^{-t\Dvec}|y>
          -\frac{1-\ga_5}{2}     <x|e^{-t\Dvec'}|y>) 
             ].
\end{eqnarray}
Let us consider the consistent and covariant anomalies.
%%%%%%%%%%%%%%%%%%%%%%%%%%%
%%%%%%%%%%%%%%%%%%%%%%%%%%%     consistent anomaly
%%%%%%%%%%%%%%%%%%%%%%%%%%%
%
%
\begin{flushleft}
(i) {\bf Consistent anomaly}
\end{flushleft}
%\subsection{consistent anomaly}

This is obtained by
setting the heat kernel operators $\Dvec$ and ${\Dvec}'$ as
%** 5.1a  %%%%%%%%%%%%%%%%%%%%%%%%%%%%%%%
\begin{eqnarray}
\Dvec = - \Dslash_{R+} \Dslash_{R+}\com\q
{\Dvec}'=\Dvec^{\dag} = - \Dslash_{R-} \Dslash_{R-}.\label{5.1a}
\end{eqnarray}
These operators are not hermitian. 
We find
%%%%%%%%%%%%%%%%%%%%%%%%%%%%%%%%%
\begin{eqnarray}
\exp(t\Dslash_{R+} \Dslash_{R+} )
&=& \exp[t \{ \dslash
(\frac{1-\gamma_5}{2}) \Dslash_R (\frac{1+\gamma_5}{2}) 
+  \Dslash_R (\frac{1+\gamma_5}{2}) \dslash (\frac{1-\gamma_5}{2}) \}]
                                                               \nn \\
&=&  \frac{1+\gamma_5}{2} \exp(t \dslash \Dslash_R )
+ \frac{1-\gamma_5}{2} \exp({t \Dslash_R \dslash })
                                                         \com \nn \\
\exp(t\Dslash_{R-} \Dslash_{R-} )
&=&  \frac{1+\gamma_5}{2} \exp(t \Dslash_R \dslash )
+ \frac{1-\gamma_5}{2} \exp({t \dslash \Dslash_R })
                                                         \com \nn \\
\ln J_{con}
&=& i \lim_{t\ra +0}{\rm Tr} \lambda(x) 
[ \gamma_5 <x| \exp(t \dslash \Dslash_R) |y>]\pr
\end{eqnarray}
The quantities $M$, $N_\mu$ and $W_{\mu\nu}$ for
the operator $-\dslash \Dslash_R$ are 
%%%%%%%%%%%%%%%%%%%%%%%%%%%%%%%%%%%%
\begin{eqnarray}
 M = -i \gamma^\mu \gamma^\nu \partial_\mu R_\nu\com\q
N_\mu = - i \gamma^\mu \gamma^\nu R_\nu\com\q
W_{\mu\nu} = 0\pr
\end{eqnarray}
The corresponding covariants $X(x)$ and $Y_{\mu\nu}(x)$ are found to be
%*** curvat %%%%%%%%%%%%%%%%%%%%%%%%%%%%%%%%%%%%%
\begin{eqnarray}
 X & = &
 -i \frac{1}{2} \partial_\mu R_\mu - \frac{1}{2} R_\mu R_\mu 
- \frac{1}{8}i [ \gamma^\mu, \gamma^\nu ] F^R_{\mu\nu}, \nn \\
Y_{\mu\nu} & = & - \half i\gamma^\nu \gamma^\lambda \partial_\mu R_\lambda 
- \frac{1}{4} \gamma^\mu \gamma^\lambda \gamma^\nu \gamma^\rho
R_\lambda R_\rho - ( \mu \leftrightarrow \nu ), \nn \\
F^R_{\mu\nu} &=& \partial_\mu R_\nu - \partial_\nu R_\mu - i[R_\mu , R_\nu].
\label{curvat}
\end{eqnarray}
Here $X$ and $Y_\mn$ are not expressed covariantly in terms
of the gauge field $R_\m$. 
(This is because the general gauge symmetry
(\ref{trawnm}),
which is introduced irrespective of the explicit forms of
($W_\mn, N_\m, M$), 
cannot be realized by the gauge transformation of $R_\m$). 
This must be compared with the covariant anomaly of the next item.
Finally the consistent anomaly
is determined to be 
%** conanom %%%%%%%%%%%%%%%%%%%%%%%%%%%%%%%%%%%%%%%%
\begin{eqnarray}
\ln J_{con} & = & i \frac{1}{(4 \pi)^2} (\frac{2}{3} )
\epsilon^{\mu\nu\lambda\rho}{\rm Tr}~\la(x)  
\left[ \partial_\mu ( R_\nu \partial_\lambda R_\rho 
- \frac{i}{2} R_\nu R_\lambda R_\rho ) \right]\com
\label{conanom}
\end{eqnarray}
where the convention 
${\rm tr}\ga_5\ga^\m \ga^\n \ga^\la \ga^\si=-4\ep^{\mn \ls},
\ep^{1234}=1$,
is used.
This coincides with the result of the conventional calculation \cite{AG}.

If we take $\Dvec$ and $\Dvec'$ in (\ref{5.1a}) in the exchanged way:\
$
\Dvec = - \Dslash_{R-} \Dslash_{R-},\ 
{\Dvec}' = - \Dslash_{R+} \Dslash_{R+}
$
,
we get 
$
\ln J_{con}
= i \lim_{t\ra +0}{\rm Tr} \lambda(x) 
[ \gamma_5 <x| \exp(t \Dslash_R \dslash) |y>]
$.
The  $-\Dslash_R \dslash$ operator is written as 
%%%%%%%%%%%%%%%%%%%%%%%%%%%%%%%%%%%%%%
\begin{eqnarray}
 M = 0\com\q
 N_\mu = - i\gamma^\nu \gamma^\mu R_\nu\com\q
 W_{\mu\nu} = 0\pr
\end{eqnarray}
The corresponding covariants, $X(x)$ and $Y_{\mu\nu}(x)$, are written as 
%%%%%%%%%%%%%%%%%%%%%%%%%%%%%%%%%%%%%%%%
\begin{eqnarray}
&& X =
 \frac{1}{2}i \partial_\mu R_\mu - \frac{1}{2} R_\mu R_\mu 
- \frac{1}{8}i [ \gamma^\mu, \gamma^\nu ] F^R_{\mu\nu}\com \nn \\
&& Y_{\mu\nu} = - \half i\gamma^\lambda \gamma^\nu \partial_\mu R_\lambda 
- \frac{1}{4} \gamma^\lambda \gamma^\mu \gamma^\rho \gamma^\nu
R_\lambda R_\rho - ( \mu \leftrightarrow \nu ).
\end{eqnarray}
The final anomaly is the same as (\ref{conanom}).

%%%%%%%%%%%%%%%%%%%%%%  covariant anomaly  %%%%%%%%%%%
\begin{flushleft}
(ii) {\bf Covariant anomaly}
\end{flushleft}
%\subsection{covariant anomaly}

The covariant anomaly is obtained by taking the
heat kernel operators  as 
%%%%%%%%%%%%%%%%%%%%%%%%%%%%%%%%%%%%%%%%
\begin{eqnarray}
\Dvec =  \Dslash_{R+}^{\dag} \Dslash_{R+} \com\q
{\Dvec}' =  \Dslash_{R+}\Dslash_{R+}^{\dag}\com
\end{eqnarray}
which are hermitian and invariant for the $R_\m$ gauge transformation.
%The differential operator $\dslash$ above is that one which
%appears in the field equation for the left-hand field.
The anomaly is evaluated as:
%%%%%%%%%%%%%%%%%%%%%%%%%%%%%%%%%%%%
\begin{eqnarray}
\exp(-t\Dslash_{R+}^{\dag} \Dslash_{R+} )
&=& \exp[t \{ \Dslash_R \Dslash_R \frac{1+\gamma_5}{2} 
             + \dslash \dslash \frac{1-\gamma_5}{2} \}]
                                                               \nn \\
&=&  \frac{1+\gamma_5}{2} \exp(t \Dslash_R \Dslash_R )
+ \frac{1-\gamma_5}{2} \exp({t \dslash \dslash })
                                                         \com \nn \\
\exp(-t\Dslash_{R+} \Dslash_{R+}^{\dag} )
&=&  \frac{1-\gamma_5}{2} \exp(t \Dslash_R \Dslash_R )
+ \frac{1+\gamma_5}{2} \exp({t \dslash \dslash })
                                                         \com \nn \\
\ln J_{cov} 
&=& i \lim_{t\ra +0} {\rm Tr} \left[\la(x) \gamma_5 <x|\exp(t \Dslash_{R}
\Dslash_R )|y> \right].
\end{eqnarray}
%
%Here we notice there appears no terms which should be
%treated by counterterms. The clear separation between renormalization
%and anomaly takes place in the covariant case.
The operator $-(\Dslash_R)^2$ is expressed by 
the following $M$, $N_\mu$ and $W_{\mu\nu}$.
%%%%%%%%%%%%%%%%%%%%%%%%%%%%%%%%%%%%%%%
\begin{eqnarray}
 M = - i\partial_\mu R_\mu - R_\mu R_\mu - \frac{i}{4} [\gamma^\mu,
\gamma^\nu] F^R_{\mu\nu}\com\q
 N_\mu = - 2 iR_\mu\com\q
 W_{\mu\nu} = 0\com
\end{eqnarray}
where $F^R_\mn$ is defined in (\ref{curvat}).
So we have covariant quantities as follows,
%%%%%%%%%%%%%%%%%%%%%%%%%%%%%%%%%%%%%%%
\begin{eqnarray}
 X = - \frac{i}{4} [\gamma^\mu, \gamma^\nu] F^R_{\mu\nu}, \com\q 
 Y_{\mu\nu} = - iF^R_{\mu\nu}\pr
\end{eqnarray}
In the covariant anomaly, all covariants ($X,Y_\mn$) under
the general gauge transformation (\ref{trawnm}) are
also covariant under the $R_\m$-gauge transformation (\ref{rmugauge}).
Substituting the above expressions into the anomaly formula
(\ref{4dformW0}), the anomaly is 
%%%%%%%%%%%%%%%%%%%%%%%%%%%%%%%%%%%%%%%%%
\begin{eqnarray}
\ln J_{cov} & = & i \frac{1}{(4 \pi)^2} (\frac{1}{2} )
\epsilon^{\mu\nu\lambda\rho}{\rm Tr}\left[ \la(x)  
F^R_{\mu\nu} F^R_{\lambda\rho} \right]\com
\end{eqnarray}
which is the covariant anomaly \cite{F79,F85}.

%%%%%%%%%%%%%%%%%%%%%%%%%%%%%%%%%%%%%%%%%%%%%%%%%%%%%%%%%%%%%%%%%%%%%
%%%%%%%%%%%%%%%%%%%%%%%%%%%%%%   SEC  5    %%%%%%%%%%%%%%%%%%%%%%%%%%
%%%%%%%%%%%%%%%%%%%%%%%%%%%%%%%%%%%%%%%%%%%%%%%%%%%%%%%%%%%%%%%%%%%%%
\section{Conclusion}

We have explained the general  gauge symmetry which appears
in the general heat-kernel $G(x,y;t)$. 
The symmetry simplifies and clarifies the structure of the anomaly
formulae. They are expressed by the covariants.
The general gauge symmetry is attributed to the local gauge
freedom of the Hilbert space on which the operator $\Dvec$ acts.
This analysis is in contrast with  the case of the counterterm formula
\cite{tH} where the symmetry is that of the action. In fact the
4-dim anomaly formula of (\ref{4dformW0}) is different from
the 't Hooft's by a total derivative term.

We  have obtained the new anomaly formulas
for two and four dimensional theories. 
These are valid for all
anomalies in most quantum field theories both on flat and 
on curved spacetimes.
Messy integral-calculation
of the anomaly has been eliminated. 
Using $W_\mn, N_\m, M$ in $\Dvec$, (\ref{cov.3}), 
(which corresponds to `which theory' we take) 
and $\Sigma$ in (\ref{GS.2}) or
(\ref{GS.7}) (which corresponds to `which symmetry' 
we consider anomoulous), 
its anomaly is expressed by (\ref{GS.11}) for
the bosonic case and by (\ref{GS.12}) for the fermionic case.
The heat-kernel $G(x,x;t)|_{t^0}$ in (\ref{GS.11}) or (\ref{GS.12})
is given by (\ref{f.47})-(\ref{4dformula})
depending on the relevent situation. 
The generalization to higher dimensional anomaly
formulae is straightforward.

In the initial condition of the heat-kernel\ 
$
\lim_{t\ra +0}~G(x,y;t)=\lim_{t\ra +0}<x|e^{-t\Dvec}|y>=\delta^n(x-y)
$
, $\del^n(x-y)$\ is a real quantity (distribution). This fact, however,
does not necessarily imply the hermiticity condition 
on $\Dvec$.\footnote{
In \cite{II}, we insufficiently explained this point.
}
In fact we have proved, in subsec.IIB of \cite{II}, the existence
of a (perturbative) solution of $G(x,y;t)$, which satisfies the
above initial condition, for a general $\Dvec$ which includes
the non-hermitian case. The assumption used in the proof is
that the weak-field perturbation is valid and the $0$-th order
of $\Dvec^{ij}$ is $-\del_\mn \del^{ij}\pl_\m \pl_\n$.  
%****** $\Dvec$ no hermiticity ****

As examples, we have calculated both consistent and covariant non-abelian
anomalies. 
The different appearance of the same anomaly in the same theory is
caused by the freedom in the choice
of the heat kernel operators $\Dvec$.
The (covariantly-expressed) anomaly formulae most efficiently work
in a covariant approach as shown in the covariant nonabelian anomaly. 
Other examples are shown in \cite{II}.

Much more needs to be explored in the anomaly.
Its most popular usage is, at present, 
the anomaly cancellation
for model building in various theories, 
including the (super)string theories (critical string).
It could have, however, a more positive meaning
as demonstrated in  2-dim models 
for the case of the chiral anomaly by Polyakov and Wiegmann\cite{PW83}
and by Witten\cite{W84} (WZNW-model)
and in the case of Weyl anomaly by Polyakov\cite{Pol87}
and by Knizhnik et al\cite{KPZ88} (non-critical string). 
For clarification, we
point out the importance of the gauge or gravitational quantum
effects to anomalies, especially to the Weyl anomaly. 
The Weyl anomaly is directly related to the renormalization-group
$\beta$-function and contains rich dynamical information of the system.
Compared with the chiral anomaly, 
the geometrical description of the Weyl anomaly is not so clear
\cite{BPB86,DS93,AKMM95,KMM96}. Furthermore  
the higher loop (higher than 1-loop) effect has not been examined.
We believe the present results are useful for  the analysis to
such a direction.

%and the Higher dimensional Weyl anomalies are not obtained.
%The anomaly formulas is useful to obtain the Weyl anomalies.
%The calculation of six dimensional Weyl anomaly is in preparation.

%Higher loop contribution is important to the Weyl anomaly and the
%higher order of the effective action. The generalization of this
%approach to the higher loop is the important problem.
%The application to quantum gravity or the supersymmetric field theory
%is interesting.

\vs 1
%%%%%%%%%%%%%%%%%%%%%%%%%%%%%%%%%%%%%%%%%%%%%%%%%%%%%%%%%%%%%%%%%%
%%%%%%%%%%%%%%%%%%%%%%%% Acknowledgement %%%%%%%%%%%%%%%%%%%%%%%%%%%%%
%%%%%%%%%%%%%%%%%%%%%%%%%%%%%%%%%%%%%%%%%%%%%%%%%%%%%%%%%%%%%%%%%%
\begin{flushleft}
{\bf Acknowledgement}
\end{flushleft}
The authors thank Prof. K.Fujikawa and 
Prof. H.Osborn for reading the manuscript carefully
and valuable comments.
They also thank 
Dr. H.Shanahan for closely checking the manuscript and
Dr. R.Kuriki 
for bringing the references
\cite{AKMM95} and \cite{KMM96} to their attention.
One(S.I.) of the authors thanks the Japanese ministry of education for
the partial financial support
(Researcher No:\ 40193445, Institution No:\ 23803).

\vs 1
%%%%%%%%%%%%%%%%%%%%%%%%%%%%%%%%%%%%%%%%%%%%%%%%%%%%%%%%%%%%%%%%%%
%%%%%%%%%%%%%%%%%%%%%%%% reference %%%%%%%%%%%%%%%%%%%%%%%%%%%%%%%
%%%%%%%%%%%%%%%%%%%%%%%%%%%%%%%%%%%%%%%%%%%%%%%%%%%%%%%%%%%%%%%%%%

\end{document}